\documentclass[a4paper,11pt]{article}
\pdfoutput=1

\usepackage{jheppub}
\usepackage[T1]{fontenc}
\usepackage{hyperref}
\usepackage{graphicx}
\usepackage{amsmath}
\usepackage{epsfig}
\usepackage{subfig}
\usepackage{xcolor}
\usepackage{natbib}
\usepackage{slashed}

\title{Fragmentation functions for gluon into $B_c$ or $B_c^*$ meson}

\author{Xu-Chang Zheng,$^{a,b}$}
\author{Chao-Hsi Chang $^{c,d,e}$ and}
\author{Xing-Gang Wu $^{a,b}$}
\affiliation{$^a$ Department of Physics, Chongqing University, \\$~~$ Chongqing 401331, P.R. China\\
$^b$ Chongqing Key Laboratory for Strongly Coupled Physics, Chongqing University, \\$~~$ Chongqing 401331, P.R. China\\
$^c$ Key Laboratory of Theoretical Physics, Institute of Theoretical Physics, \\$~~$ Chinese Academy of Sciences, Beijing 100190, P.R. China. \\
$^d$ School of Physical Sciences, University of Chinese Academy of Sciences, \\$~~$ Beijing 100049, P.R. China.\\
$^e$ CCAST (World Laboratory), \\$~~$ Beijing 100190, P.R. China}
\emailAdd{zhengxc@cqu.edu.cn, zhangzx@itp.ac.cn, wuxg@cqu.edu.cn}

\abstract{In the paper, we calculate the fragmentation functions for $g \to B_c$ and $g \to B_c^*$. The ultraviolet divergences in the calculation are removed through the renormalization of the operator definition of the fragmentation functions under the modified minimal subtraction scheme. We then obtain the fragmentation functions $D_{g \to B_c}(z,\mu_F)$ and $D_{g \to B_c^*}(z,\mu_F)$, which are presented as figures and fitting functions. The obtained fragmentation functions are complementary to the previous work on the next-to-leading order fragmentation functions for $\bar{b}\to B_c(B_c^*)$ and $c\to B_c(B_c^*)$.
}


\begin{document}

\maketitle

\bibliographystyle{JHEP}

\section{Introduction}
\label{intro}

The $B_c$ meson has attracted a great deal of attention since its first observation by the CDF collaboration at the Tevatron. The $B_c$ meson is very interesting because it is the only meson in the Standard Model (SM) with two different heavy flavors. Its production differs significantly from that of charmonium and bottomonium as well as from that of hadrons containing only one heavy quark. Unlike the production of the hadrons containing one heavy quark, a lot of information for the $B_c$ meson production can be calculated reliably through perturbative QCD (pQCD) theory. In flavor conserving collisions, there are $c\bar{c}$ and $b\bar{b}$ pairs should be created simultaneously for the production of the $B_c$ meson, which makes the production of the $B_c$ meson be more difficult than that of charmonium and bottomonium. Therefore, the $B_c$ meson production provides a special platform for studying the strong interactions among quarks and gluons.

The production of the $B_c$ meson at large transverse momentum ($p_T$) region is simpler than other cases since the long-distance interactions between the $B_c$ and initial particles are suppressed. Thus, it is important to study the production of the $B_c$ meson at large transverse momentum region for understanding the production mechanisms of the $B_c$ meson. According to QCD factorization theorem, the production cross section for a hadron at large $p_T$ region is dominated by the fragmentation mechanism \cite{Collins:1989gx}, i.e.,
\begin{eqnarray}
d\sigma_{A+B \to H(p_T)+X}=&& \sum_i d \hat{\sigma}_{A+B\to i+X}(p_T/z,\mu_F) \otimes D_{i\to H}(z,\mu_F)+{\cal O}(m_H^2/p_T^2),
\label{pqcd-fact}
\end{eqnarray}
where $d \hat{\sigma}_{A+B\to i+X}(p_T/z,\mu_F)$ are partonic production cross sections, $D_{i\to H}(z,\mu_F)$ are fragmentation functions for a parton into the hadron $H$, $\mu_F$ is the factorization scale, and $\otimes$ denotes the convolution integral over $z$.

Fragmentation functions $D_{i\to H}(z,\mu_F)$ play a central role in the factorization formulism (\ref{pqcd-fact}). Unlike the fragmentation functions for light hadrons which are nonperturbative in nature, the fragmentation functions for the $B_c$ meson can be calculated through the nonrelativistic QCD (NRQCD) factorization \cite{nrqcd}, i.e.,
\begin{eqnarray}
D_{i\to B_c}(z,\mu_F)=\sum_n d_{i\to (c\bar{b})[n]}(z,\mu_F) \langle {\cal O}^{B_c}(n) \rangle,
\label{frag-nrqcd}
\end{eqnarray}
where $d_{i\to (c\bar{b})[n]}(z,\mu_F)$ are short-distance coefficients (SDCs) which can be calculated perturbatively, $\langle {\cal O}^{B_c}(n) \rangle$ are long-distance matrix elements (LDMEs) which are nonperturbative but can be determined by fitting experimental data or estimated by the QCD potential models.

The leading order (LO) fragmentation functions for $Q \to B_c(B_c^*)$, where $Q=\bar{b}$ or $c$, were first correctly calculated by the authors of ref.\cite{Chang:1992bb}. They extracted the LO fragmentation functions from the LO calculation of the processes $Z \to B_c(B_c^*)+ b+\bar{c}$ by taking the limit $m_{Bc}/m_{_Z} \to 0$. The subsequent calculations using different methods by other groups confirmed their results \cite{Braaten:1993jn, Ma:1994zt}. The LO fragmentation functions for the $P$-wave and $D$-wave excited states of the $B_c$ meson were calculated in refs.\cite{Chen:1993ii, Yuan:1994hn, Cheung:1995ir}. In recent years, with the development of the loop-diagram calculation techniques, some fragmentation functions for the heavy quarkonia have been calculated to the higher order of $\alpha_s$ and $v$ \cite{Braaten:2000pc, Artoisenet:2014lpa,  Artoisenet:2018dbs, Feng:2018ulg, Zhang:2018mlo, Zheng:2019dfk, Zheng:2019gnb, Yang:2019gga, Feng:2017cjk, Feng:2021uct, Zhang:2020atv, Chen:2021hzo, Zheng:2021mqr, Zheng:2021ylc}, where $v$ is the relative velocity of the constituent quarks in the quarkonia rest frame. Among those studies, the NLO corrections to the fragmentation functions $D_{Q \to B_c(B_c^*)}(z,\mu_F)$, where $Q=\bar{b}$ or $c$, have been obtained in our previous work \cite{Zheng:2019gnb}. However, the fragmentation functions for $g \to B_c(B_c^*)$, which start at order $\alpha_s^3$, are absent now. These gluon fragmentation functions are also important to the precision prediction of the $B_c$ production cross section at large $p_T$ region. In this paper, we devote ourselves to calculating the fragmentation functions $D_{g \to B_c}(z,\mu_F)$ and  $D_{g \to B_c^*}(z,\mu_F)$.

In ref.\cite{Cheung:1993pk}, Cheung and Yuan calculated the gluon fragmentation functions for the $B_c$ and $B_c^*$ production through solving the Dokshitzer-Gribov-Lipatov-Altarelli-Parisi (DGLAP) evolution equations, and they found that the gluon fragmentation can give a significant contribution to the production of the $B_c(B_c^*)$ at the Tevatron. For solving the DGLAP equations, the fragmentation functions $D_{Q \to B_c(B_c^*)}(z, \mu_{F0})$ and $D_{g \to B_c(B_c^*)}(z,\mu_{F0})$ at an initial factorization scale $\mu_{F0}={\cal O}(m_Q)$ should be input as the boundary conditions. However, the authors of ref.\cite{Cheung:1993pk} simply set the initial fragmentation functions $D_{g \to B_c(B_c^*)}(z,\mu_{F0})$ to zeros, e.g. $D_{g \to B_c(B_c^*)}(z,\mu_{F0}<2m_{b} +2m_{c})=0$. Thus, the gluon fragmentation functions obtained in ref.\cite{Cheung:1993pk} are incomplete. In this paper, we will calculate the complete gluon fragmentation functions for the $B_c$ and $B_c^*$ production at order $\alpha_s^3$.


The paper is organized as follows. In Sec.\ref{calcmethod}, we present the definition of fragmentation function and sketch the method used in the calculation of the fragmentation functions. In Sec.\ref{numer}, we present the numerical results for the fragmentation functions $D_{g \to B_c}(z,\mu_F)$ and $D_{g \to B_c^*}(z,\mu_F)$. Section \ref{sum} is reserved for a summary.

\section{Calculation method for the fragmentation function}
\label{calcmethod}

\subsection{Definition of fragmentation function}

In this section, we will sketch the method used to calculate the fragmentation function. Before carrying out the calculation, we first present the definition of the fragmentation function. We adopt the gauge-invariant definition of the fragmentation function given by Collins and Soper in ref.\cite{Collins:1981uw}. For a gluon fragmenting into a hadron, the fragmentation function is defined as
\begin{eqnarray}
D_{g\to H}(z)=&&\frac{-g_{\mu \nu}\,z^{d-3}}{2\pi K^+ (N_c^2-1)(d-2)}\sum_{X} \int dx^- e^{-iP^+ x^-/z} \langle 0 \vert G_c^{+\mu}(0) \mathcal{E}^{\dagger}(0,0,\textbf{0}_{\perp})_{cb}\nonumber \\
&&\times  \vert H(P^+,{\bf 0}_\perp)+X \rangle \langle H(P^+,{\bf 0}_\perp)+X\vert \mathcal{E}(0,x^-,\textbf{0}_{\perp})_{ba} G_a^{+\nu}(0,x^-,\textbf{0}_{\perp}) \vert 0\rangle.
\label{defrag1}
\end{eqnarray}
Here, the light-cone coordinates are used to define the fragmentation function, where a d-dimensional vector is expressed as $V^{\mu}=(V^+,V^-,\textbf{V}_{\perp})=((V^0+V^{d-1})/\sqrt{2},(V^0-V^{d-1})/\sqrt{2},\textbf{V}_{\perp})$ and the product of two vectors becomes $V \cdot W=V^+ W^- +V^- W^+ -\textbf{V}_{\perp}\cdot \textbf{W}_{\perp}$. $G^{\mu \nu}$ is the gluon field-strength operator, $K$ is the momentum of the initial gluon, $P$ is the momentum of the final hadron $H$, $z\equiv P^+/K^+$ is the  longitudinal momentum fraction. The gauge link $\mathcal{E}(0,x^-,\textbf{0}_{\perp})_{ba}$ is
\begin{eqnarray}
\mathcal{E}(0,x^-,\textbf{0}_{\perp})_{ba}={\cal P}{\rm exp}\left[ ig_s \int_{x^-}^{\infty}dz^- A^+(0,z^-,\textbf{0}_{\perp})  \right]_{ba},
\end{eqnarray}
where ${\cal P}$ denotes the path ordering, $[A^{\mu}(x)]_{ba}=-if^{cba}A^{\mu}_c(x)$ is the matrix-valued gluon
field in the adjoint representation.

The definition (\ref{defrag1}) is carried out in a reference frame in which the transverse momentum of the produced hadron $H$ vanishes. It is convenient to introduce a lightlike vector which has the components as $n^{\mu}=(0,1,\textbf{0}_{\perp})$ in the reference frame where the fragmentation function is defined. Then, the ``+" component of a momentum $p$ can be expressed as a Lorentz invariant, i.e., $p^+=p\cdot n$.

According to the definition in Eq.(\ref{defrag1}), the Feynman rules for the gluon fragmentation function can be derived directly. The QCD Feynman rules are also applicable here, and the additional Feynman rules specified to the gluon fragmentation function can be summarized as follows. The fragmentation function is expressed as the sum of cut diagrams, and each cut diagram contains an eikonal line (gauge link) connecting the operators on both sides of the cut. There is an overall factor
\begin{eqnarray}
N_{\rm CS}=\frac{1}{(N_c^2-1)(d-2)}\frac{z^{d-3}}{2\pi K\cdot n},
\label{eq.Ncs}
\end{eqnarray}
which arises from the definition. For the Feynman rules relating to the eikonal lines, we only state the rules for the left part to the cut line. The rules for the right part to the cut line can be obtained by complex conjugation. The operator vertex that creates a gluon and an eikonal line contributes a factor
\begin{eqnarray}
-i(K\cdot n \,g^{\mu \lambda}-q^{\mu}n^{\lambda})\delta_{ac},
\end{eqnarray}
where $K$ is the sum of the momenta of the created gluon and the eikonal line, $q$ is the momentum of the created gluon, $\lambda$ is the Lorentz index of the created gluon, $a$ and $c$ are the color indices of the eikonal line and the created gluon. The eikonal-line-gluon vertex contributes a factor $g_s n^{\mu}f^{abc}$, where $\mu$ and $a$ are the Lorentz and color indices of the gluon, $b$ and $c$ are the left and right color indices of the eikonal line. The propagator for an eikonal line carrying momentum $l$ contributes a factor $i\delta_{ab}/(l\cdot n+i \epsilon)$. The cut through an eikonal line carrying momentum $l$ contributes a factor $2\pi \delta(l\cdot n)$.

\subsection{Calculation technology}

\begin{figure}[htbp]
\centering
\includegraphics[width=0.6\textwidth]{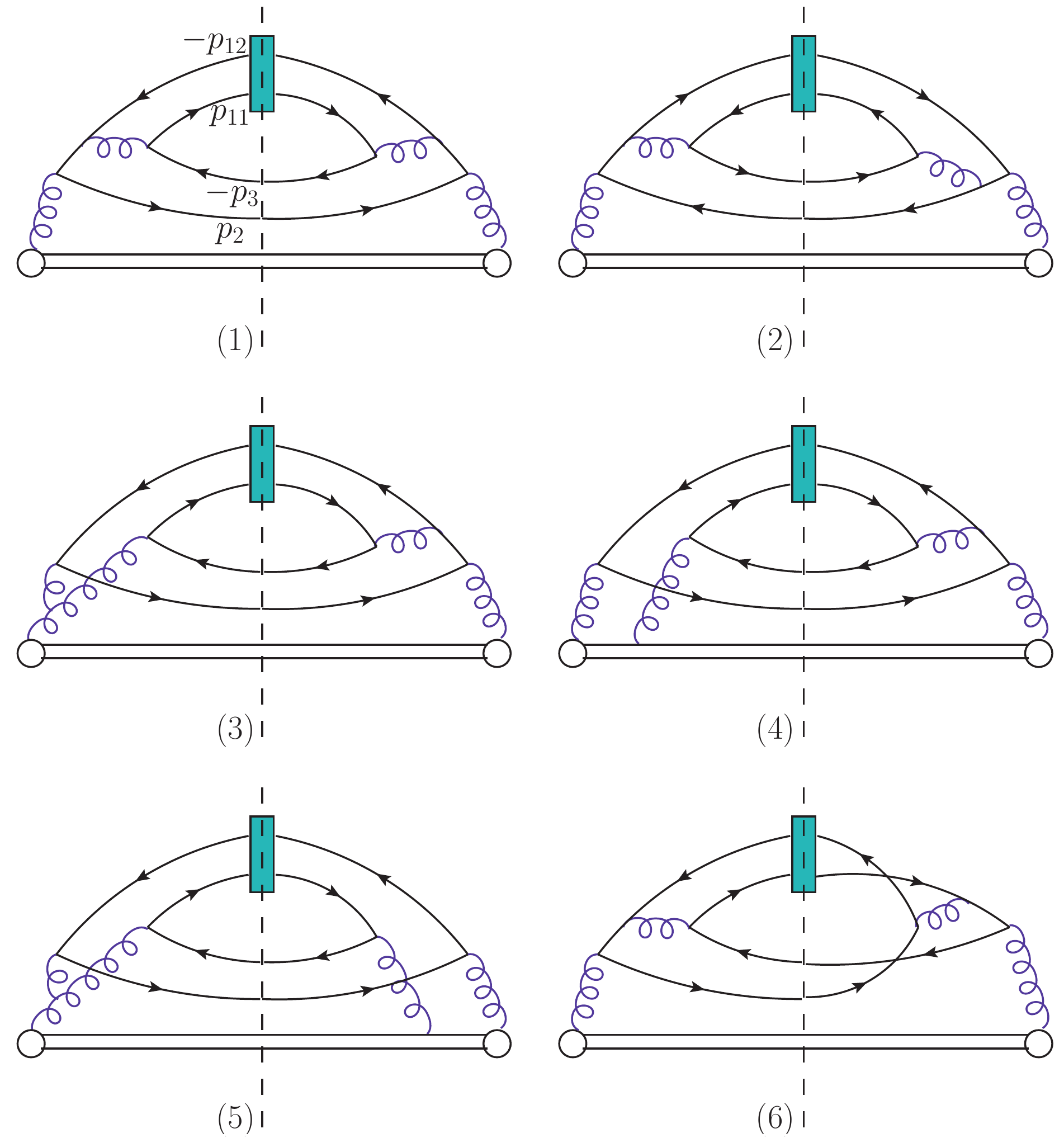}
\caption{Six of 49 cut diagrams for the fragmentation functions $D_{g \to (c\bar{b})[n]}(z,\mu_F)$.
 } \label{feyn}
\end{figure}

For simplicity, in the rest of this section, we only present the formulas for the fragmentation function $D_{g \to B_c}(z,\mu_F)$. The formulas for the fragmentation function $D_{g \to B_c^*}(z,\mu_F)$ are similar.

In the practical calculation, we first calculate the fragmentation function for producing a free $(c\bar{b})[n]$ pair with $n= \,^1S_0^{[1]}$. Then the fragmentation function for the $B_c$ meson can be obtained from $D_{g \to (c\bar{b})[^1S_0^{[1]}]}(z,\mu_F)$ through the replacement of $\langle {\cal O} ^{c\bar{b}[^1S_0^{[1]}]}(^1S_0^{[1]})\rangle \to \langle {\cal O} ^{B_c}(^1S_0^{[1]})\rangle$.

There are totally 49 cut diagrams for the fragmentation function $D_{g \to (c\bar{b})[^1S_0^{[1]}]}(z,\mu_F)$ under the Feynman gauge, six of which are shown in Fig.\ref{feyn}. According to the Feynman rules, the squared amplitude for the fragmentation function can be written down directly. In the calculation, the package FeynCalc \cite{Mertig:1990an,Shtabovenko:2016sxi} is employed to carry out the Dirac and color traces.

To obtain the amplitude for a free $(c\bar{b})$ pair in $^1S_0^{[1]}$ state, we adopt the covariant projector technique \cite{Petrelli:1997ge}. The projector for the spin singlet is
\begin{eqnarray}
\Pi_1= \frac{-\sqrt{M}}{{4{m_b}{m_c}}}(\slashed{p}_{12}- m_b) \gamma_5 (\slashed{p}_{11} + m_c).
\end{eqnarray}
For the fragmentation function of $g\to B_c^*$, the projector for the spin triplet is needed, and its expression is
\begin{eqnarray}
\Pi_3= \frac{-\sqrt{M}}{{4{m_b}{m_c}}}(\slashed{p}_{12}- m_b) \slashed{\epsilon}(p_1) (\slashed{p}_{11} + m_c),
\end{eqnarray}
where $M=m_b+m_c$, $r_c=(1-r_b)=m_c/M$, $p_{11}=r_c\,p_1$, $p_{12}=r_b\,p_1$, and $p_1$ is the momentum of the produced $(c\bar{b})[n]$ pair. The color projector for the color singlet is
\begin{eqnarray}
\Lambda_1= \frac{1}{\sqrt{3}} \textbf{1},
\end{eqnarray}
where $\textbf{1}$ denotes the unit matrix of the $SU(3)$ group.

Having the squared amplitude, the total contribution from the 49 cut diagrams can be calculated through
\begin{eqnarray}
D^{(1-49)}_{g \to (c\bar{b})[^1S_0^{[1]}]}(z)=N_{\rm CS}\int d\phi_{3}(p_1,p_2,p_3) {\cal A}_{(1-49)},
\label{cut-contribution}
\end{eqnarray}
where $N_{\rm CS}$ is the overall factor whose expression has been given in Eq.(\ref{eq.Ncs}), ${\cal A}_{(1-49)}$ is the squared amplitude for the 49 cut diagrams, and $ d\phi_{3}(p_1,p_2,p_3)$ is the differential phase space whose expression is
\begin{eqnarray}
d\phi_{3}(p_1,p_2,p_3)=2\pi \delta\left(K^+ - \sum_{i=1}^{3}  p_i^+\right)\mu^{2(4-d)}  \prod_{i=2,3}\frac{\theta(p_i^+)dp_i^+}{4 \pi p_i^+}\frac{ d^{d-2}\textbf{p}_{i\perp}}{(2\pi)^{d-2}}.
\end{eqnarray}
The integral in Eq.(\ref{cut-contribution}) is ultraviolet (UV) divergent when $d=4$. The UV divergences of this integral arise from the phase-space regions of $\vert \textbf{p}_{2\perp} \vert \to \infty$ and $\vert \textbf{p}_{3\perp} \vert \to \infty$. We adopt the dimensional regularization with $d=4-2\epsilon$ to regularize these UV divergences. Then the UV divergences appear as pole terms in $\epsilon$.

It is impractical to calculate the integral in Eq.(\ref{cut-contribution}) in $d$ dimensions directly due to the fact that ${\cal A}_{(1-49)}$ is complicated. To carry out the integral, we adopt the subtraction method which was recently used to calculate the real corrections to fragmentation functions for quarkonia \cite{Artoisenet:2014lpa,Artoisenet:2018dbs,Zheng:2019dfk,Zheng:2019gnb,Zheng:2021mqr,Zheng:2021ylc}. Under the subtraction method, the contribution of the 49 cut diagrams can be calculated through
\begin{eqnarray}
D^{(1-49)}_{g \to (c\bar{b})[^1S_0^{[1]}]}(z)=&& N_{\rm CS}\int d\phi_{3}(p_1,p_2,p_3) \left[{\cal A}_{(1-49)}-{\cal A}_{S}\right]\nonumber \\
&&+ N_{\rm CS}\int d\phi_{3}(p_1,p_2,p_3) {\cal A}_{S},
\label{sub-method}
\end{eqnarray}
where ${\cal A}_{S}$ denotes the constructed subtraction term which has the same singularity behavior as ${\cal A}_{(1-49)}$. The first integral on the right-hand side of Eq.(\ref{sub-method}) is finite and we calculate it numerically in 4 dimensions. The second integral on the right-hand side of Eq.(\ref{sub-method}) contains the same divergences as the the integral in Eq.(\ref{cut-contribution}), and it should be calculated analytically in $d$ dimensions.

The squared amplitude for the 49 cut diagrams can be expressed as:
\begin{eqnarray}
{\cal A}_{(1-49)}=&&\frac{b_1(s_1,z,y)}{s}+\frac{b_2(s_1,z,y)p_1 \cdot p_3}{s^2}+\frac{b_3(s_1,z,y)}{s_4-m_c^2}+\frac{b_4(s_1,z,y)p_1 \cdot p_3}{(s_4-m_c^2)^2}+\frac{b_5(s_1,z,y)}{s_2-m_b^2}\nonumber \\
&& +\frac{c_1(s_2,z,y)}{s}+\frac{c_2(s_2,z,y)p_1\cdot p_2}{s^2}+\frac{c_3(s_2,z,y)}{s_3-m_b^2}+\frac{c_4(s_2,z,y)p_1\cdot p_2}{(s_3-m_b^2)^2}+\frac{c_5(s_2,z,y)}{s_1-m_c^2}\nonumber \\
&&+\frac{d(z,y)p_2\cdot p_3}{(s_1-m_c^2)^2(s_2-m_b^2)^2}+{\cal A}^{\rm finite}_{\rm real},
\end{eqnarray}
where the Lorentz-invariant quantities are defined as follows
\begin{eqnarray}
&&s=(p_1+p_2+p_3)^2,\; s_1=(p_1+p_2)^2,\nonumber \\
&&s_2=(p_1+p_3)^2, \;s_3=(p_{11}+p_2+p_3)^2,\nonumber \\
&&s_4=(p_{12}+p_2+p_3)^2,\, y=\frac{(p_1+p_2)\cdot n}{(p_1+p_2+p_3) \cdot n},\label{defvar}
\end{eqnarray}
${\cal A}^{\rm finite}_{\rm real}$ denotes the remaining terms which do not contribute divergences. The coefficients $b_i(s_1,z,y)$ behave as $1/s_1^2$ when $s_1 \to \infty$, and the coefficients $c_i(s_2,z,y)$ behave as $1/s_2^2$ when $s_2 \to \infty$.

The subtraction term can be constructed as follows:
\begin{eqnarray}
{\cal A}_{S}=&&\frac{b_1(s_1,z,y)}{s}+\frac{b_2(s_1,z,y)(p_1 \cdot p_3-f_1)}{s^2}+\frac{b_3(s_1,z,y)}{s_4-m_c^2}+\frac{b_4(s_1,z,y)(p_1 \cdot p_3-f_2)}{(s_4-m_c^2)^2}\nonumber \\
&&+\frac{b_5(s_1,z,y)}{s_2-m_b^2}+\frac{c_1(s_2,z,y)}{s}+\frac{c_2(s_2,z,y)(p_1\cdot p_2-g_1)}{s^2}+\frac{c_3(s_2,z,y)}{s_3-m_b^2} \nonumber \\
&&+\frac{c_4(s_2,z,y)(p_1\cdot p_2-g_2)}{(s_3-m_b^2)^2} +\frac{c_5'(s_2,z,y)}{s_1-m_c^2}+\frac{d(z,y)(p_2\cdot p_3-f_3)}{(s_1-m_c^2)^2(s_2-m_b^2)^2},\label{subterm}
\end{eqnarray}
where
\begin{eqnarray}
&& f_1=\frac{-[y(y-z-1)+2z]s_1+[r_c (y-z-1)-2 y+2]y\,r_c M^2}{2y^2},\label{subterm-f1}\\
&& f_2=\frac{-[z (r_c^2 z-r_c (z+1)+2)+y^2-y (z+1)](s_1-m_c^2)}{2 (y-r_c z)^2},\\
&& f_3=\frac{(1-y)z\,s_1+[r_c z (r_c (y-1)-4 y+2)+2 r_c z^2+2 (y-1) y]M^2}{2z^2},\\
&& g_1=\frac{[y (-y+z+1)-2 z]s_2-(y-z-1)(r_c y+y-2 z)r_b M^2}{2(1-y+z)^2},\\
&& g_2=\frac{-[z ( -r_c r_b z+r_c+1)+y^2-y (z+1)](s_2-m_b^2)}{2(r_c z-y+1)^2},\\
&& c_5'(s_2,z,y)=c_5(s_2,z,y)+\frac{d(z,y)(1-y)z}{2z^2(s_2-m_b^2)^2}.
\end{eqnarray}
The terms $f_i$, $g_i$, and $c_5'$ are constructed to make the integral of the subtraction term simple. An explicit example (for $f_1$) of constituting such terms can be found in Eqs.(\ref{eqb6})-(\ref{eqb11}).

To perform the integration of the subtraction term, the phase space should be parametrized properly. The phase-space parametrizations for different terms in Eq.(\ref{subterm}) are given in Appendix \ref{Ap.para}. Moreover, the integrations of ${\cal A}_S$ over the phase space under these parametrizations are presented in Appendix \ref{Ap.int}.

\subsection{Renormalization}

The UV divergences in $D^{(1-49)}_{g \to (c\bar{b})[^1S_0^{[1]}]}(z)$ should be removed through the renormalization of the operator defining the fragmentation function \cite{Mueller:1978xu}. We carry out the renormalization under the modified-minimal-subtraction scheme ($\overline{\rm MS}$). Then the fragmentation function can be obtained through
\begin{eqnarray}
&& D_{g \to (c\bar{b})[^1S_0^{[1]}]}(z,\mu_F)\nonumber\\
&&=D^{(1-49)}_{g \to (c\bar{b})[^1S_0^{[1]}]}(z)-\frac{\alpha_s}{2\pi}\left[\frac{1}{\epsilon_{_{UV}}}- \gamma_E + {\rm ln}\frac{4\pi \mu^2}{\mu_F^2} \right]\nonumber \\
&& ~~~ \times \int_z^1 \frac{dy}{y}\Bigg[\sum_{Q=\bar{b},c}P_{Qg}(y) D_{Q\to (c\bar{b})[^1S_0^{[1]}]}^{\rm LO}(z/y)\Bigg],
\label{FFRen}
\end{eqnarray}
where $D_{\bar{b}\to (c\bar{b})[^1S_0^{[1]}]}^{\rm LO}(z)$ and $D_{c\to (c\bar{b})[^1S_0^{[1]}]}^{\rm LO}(z)$ are the LO fragmentation functions in $d$-dimensional space-time. The splitting function for $g\to Q$ is
\begin{equation}
P_{Qg}(y)=T_F\left[y^2+(1-y)^2\right],
\label{lospfun}
\end{equation}
where $T_F=1/2$.

\section{Numerical results and discussion}
\label{numer}

After the renormalization, we obtain the finite results for the fragmentation functions $D_{g \to (c\bar{b})[^1S_0^{[1]}]}(z,\mu_F)$ and $D_{g \to (c\bar{b})[^3S_1^{[1]}]}(z,\mu_F)$. The fragmentation functions $D_{g \to B_c}(z,\mu_F)$ and $D_{g \to B_c^*}(z,\mu_F)$ can be obtained from $D_{g \to (c\bar{b})[n]}(z,\mu_F)$ by multiplying a factor $\langle {\cal O}^{B_c(B_c^*)}(n)\rangle  \\ / \langle {\cal O}^{c\bar{b}[n]}(n)\rangle \approx \vert R_S(0) \vert^2/4\pi$, where $n=\, ^1S_0$ or $^3S_1$ accordingly. Here, $R_S(0)$ is the radial wave function at the origin for the $(c\bar{b})$ bound system. In the calculation, we use the program Vegas \cite{Lepage:1977sw} to perform the numerical integrations.

The input parameters for the numerical calculation are taken as follows:
\begin{eqnarray}
&& m_c=1.5 ~{\rm GeV}\,,\; m_b=4.9 ~{\rm GeV}\,,\;\vert R_S(0)\vert^2=1.642~{\rm GeV^3}.
\end{eqnarray}
where the value of $\vert R_S(0)\vert^2$ is taken from the potential model calculation \cite{Eichten:1995ch}. For the strong coupling constant, we use the two-loop formula as that adopted in ref.\cite{Zheng:2019gnb}, where $\alpha_s(2m_c)=0.259$.

\begin{figure}[htbp]
\centering
\includegraphics[width=0.8\textwidth]{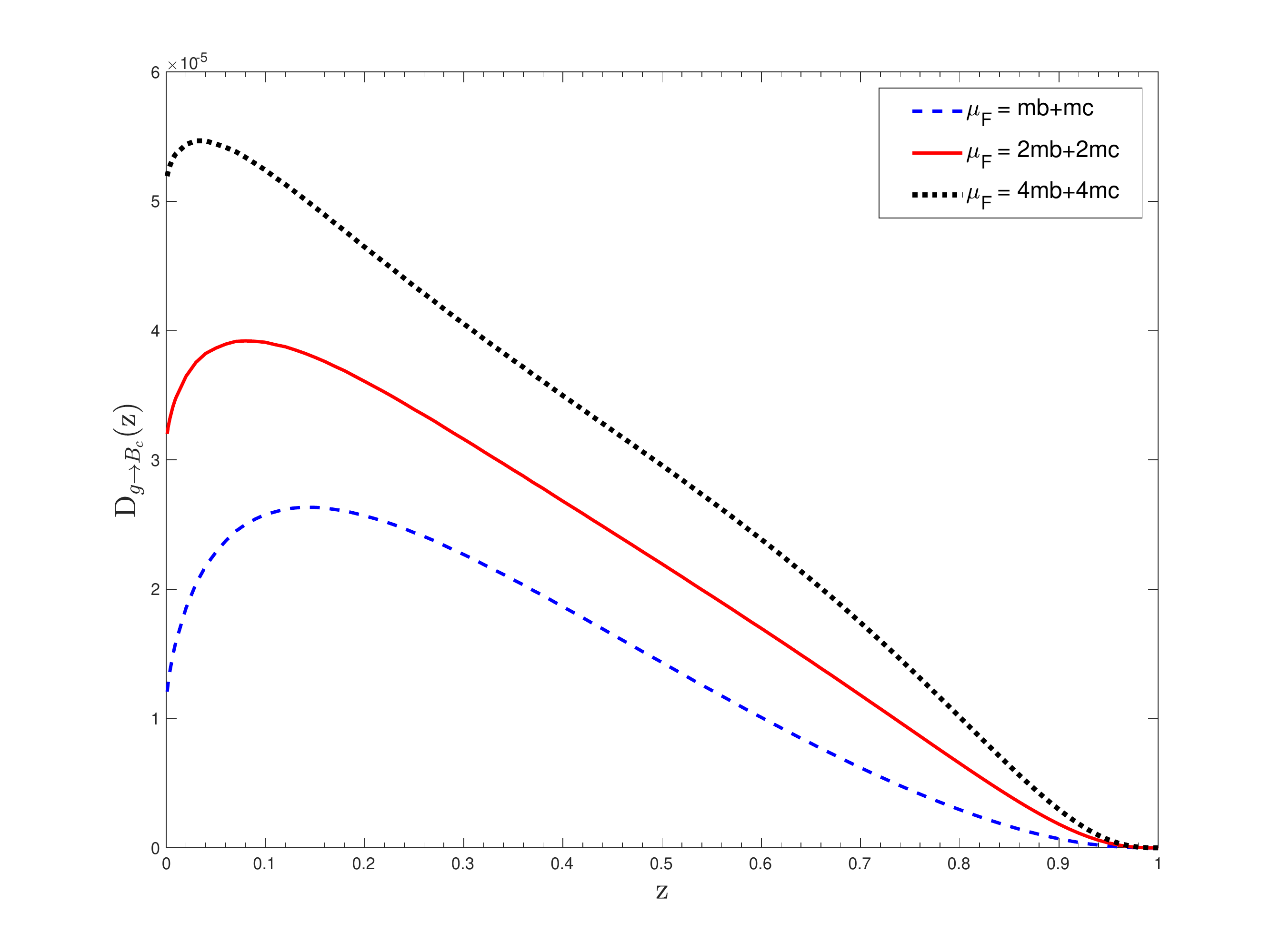}
\caption{The fragmentation function $D_{g \to B_c}(z,\mu_F)$ as a function of $z$ for $\mu_F=M$, $\mu_F=2M$, and $\mu_F=4M$, where the strong coupling constant is fixed as $\alpha_s(2m_c)=0.259$.} \label{BcFF}
\end{figure}

\begin{figure}[htbp]
\centering
\includegraphics[width=0.8\textwidth]{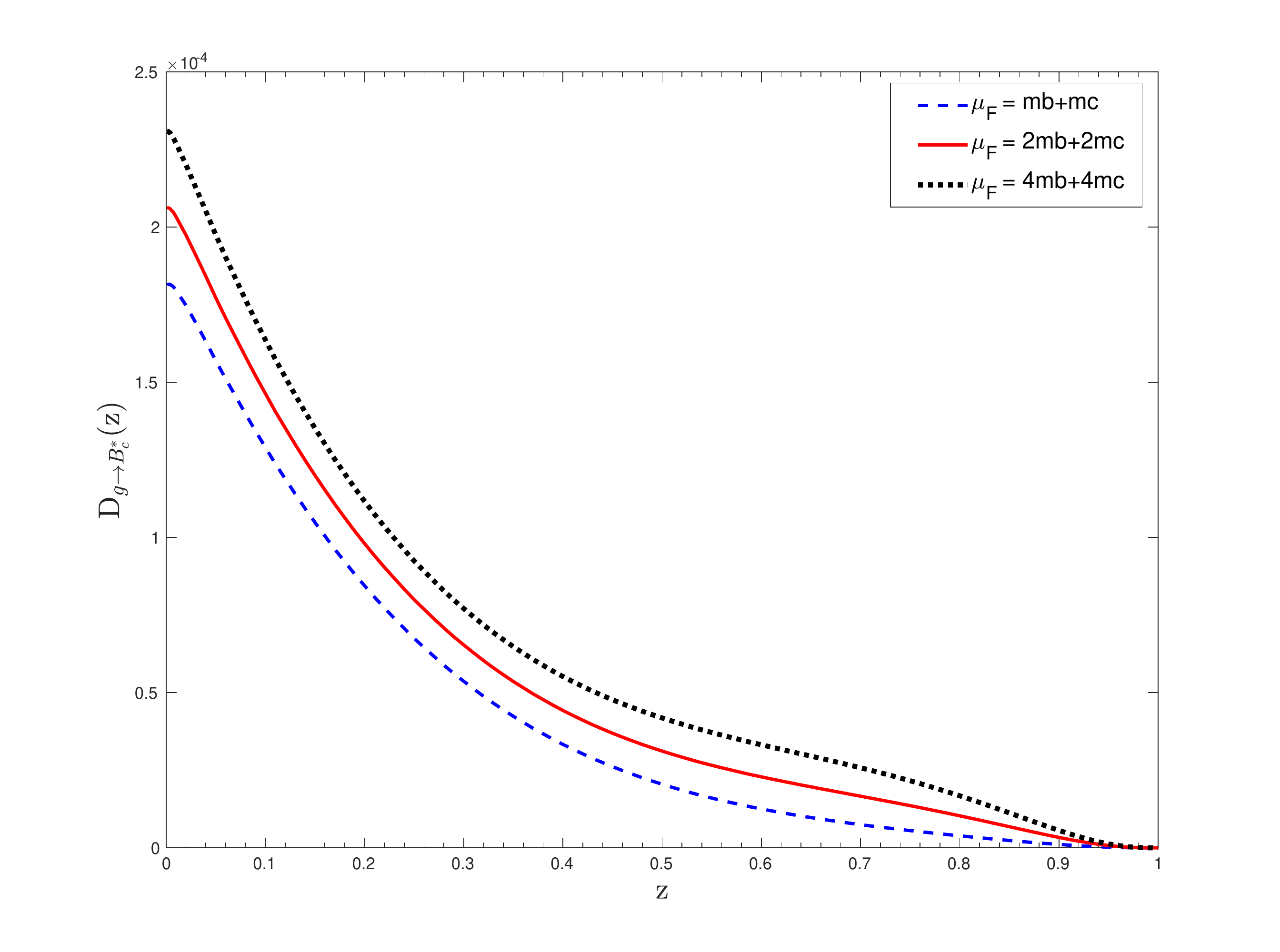}
\caption{The fragmentation function $D_{g \to B_c^*}(z,\mu_F)$ as a function of $z$ for $\mu_F=M$, $\mu_F=2M$, and $\mu_F=4M$, where the strong coupling constant is fixed as $\alpha_s(2m_c)=0.259$.
 } \label{Bc*FF}
\end{figure}

We present the fragmentation functions $D_{g \to B_c}(z,\mu_F)$ and $D_{g \to B_c^*}(z,\mu_F)$ for $\mu_F=M$ ($M=m_b+m_c$), $\mu_F=2M$, and $\mu_F=4M$ in Figs.\ref{BcFF} and \ref{Bc*FF}. For $g \to B_c$, the fragmentation function has a peak at a small $z$ value. As the $z$ value increases, the fragmentation function first increases rapidly to the maximum value, and then decreases slowly to zero. For $g \to B_c^*$, the fragmentation function has the maximum value at $z=0$. The shape of the fragmentation function for $g\to B_c(B_c^*)$ is obviously different from that of the fragmentation functions for $\bar{b}\to B_c(B_c^*)$ and $c\to B_c(B_c^*)$ \cite{Zheng:2019gnb}. The fragmentation function for $\bar{b}\to B_c(B_c^*)$ has a peak at a large $z$ value, while the fragmentation function for $c\to B_c(B_c^*)$ has a peak at an intermediate $z$ value.

\begin{figure}[htbp]
\centering
\includegraphics[width=0.8\textwidth]{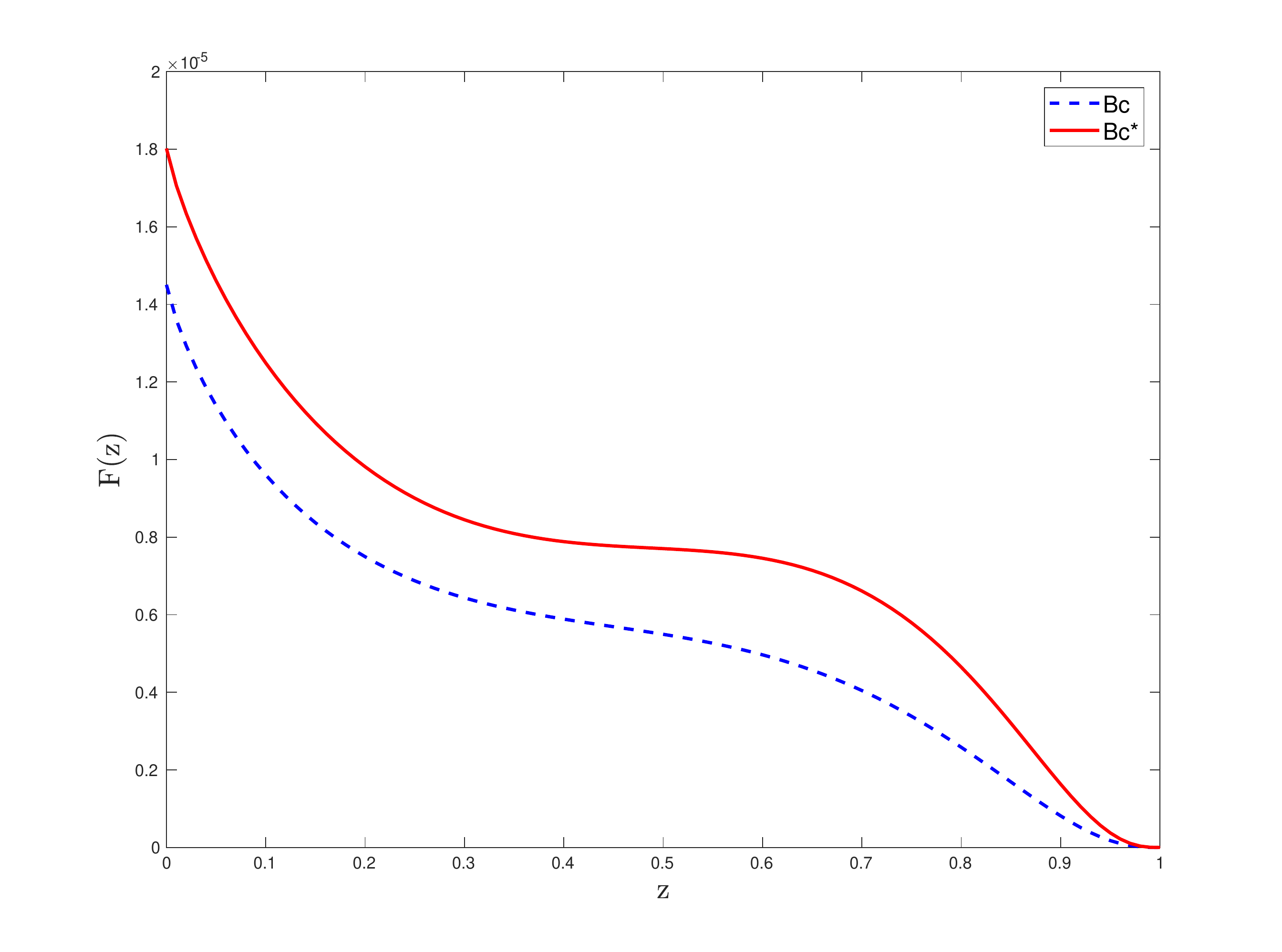}
\caption{The coefficients of ${\rm ln}(\mu_F^2/M^2)$ in the fragmentation functions $D_{g \to B_c}(z,\mu_F)$ and $D_{g \to B_c^*}(z,\mu_F)$, i.e., $F(z)=\frac{\alpha_s}{2\pi}\int_z^1 \frac{dy}{y}\big[\sum_{Q=\bar{b},c}P_{Qg}(y)D_{Q \to B_c(B_c^*)}^{\rm LO}(z/y)\big]$.
 } \label{coeff}
\end{figure}

From Figs.\ref{BcFF} and \ref{Bc*FF}, we can also see that the fragmentation functions $D_{g \to B_c}(z,\mu_F)$ and $D_{g \to B_c^*}(z,\mu_F)$ are sensitive to the factorization scale. When the factorization scale increases, the fragmentation functions increase for $z \in (0,1)$. To understand the dependence behavior of the fragmentation functions on the factorization scale, we show the coefficients of ${\rm ln}(\mu_F^2/M^2)$ in $D_{g \to B_c}(z,\mu_F)$ and $D_{g \to B_c^*}(z,\mu_F)$ as functions of $z$ in Fig.\ref{coeff}. We can see that the coefficients are positive for $z \in (0,1)$. Therefore, the fragmentation functions for $g \to B_c$ and $g \to B^*_c$ become very important for the $B_c$ and $B_c^*$ production when the energy scale involved in the process is very large.

\begin{table}[htb]
\centering
\begin{tabular}{c c c }
\hline\hline
$\mu_F$ &  ${\rm P}\times 10^{5}$~ & ~$\langle z \rangle$~ \\
\hline
$M$ & 1.36  &   0.31   \\
$2M$ & 2.10 &  0.32   \\
$4M$ & 2.85 &   0.33  \\
\hline\hline
\end{tabular}
\caption{The fragmentation probability and average $z$ value for $D_{g\to B_c}(z,\mu_{F})$ with three typical factorization scales, where the strong coupling constant is fixed as $\alpha_s(2m_c)=0.259$.}
\label{tbcfp}
\end{table}

\begin{table}[htb]
\centering
\begin{tabular}{c c c }
\hline\hline
$\mu_F$ &  ${\rm P}\times 10^{5}$~ & ~$\langle z \rangle$~ \\
\hline
$M$ & 4.33  &   0.20   \\
$2M$ & 5.37 &  0.23   \\
$4M$ & 6.40 &  0.25   \\
\hline\hline
\end{tabular}
\caption{The fragmentation probability and average $z$ value for $D_{g\to B_c^*}(z,\mu_{F})$ with three typical factorization scales, where the strong coupling constant is fixed as $\alpha_s(2m_c)=0.259$.}
\label{tbc*fp}
\end{table}

Two useful quantities can be derived from the obtained fragmentation functions, i.e., the fragmentation probability $P$ and the average $z$ value $\langle z \rangle$, which are defined as
\begin{eqnarray}
P &=& \int_0^1 D(z) dz, \\
\langle z \rangle &=& \frac{\int_0^1 z\,D(z) dz}{\int_0^1 D(z) dz},
\end{eqnarray}
where $D(z)$ denotes the fragmentation function for $g \to B_c$ or $g \to B_c^*$. The numerical results for the fragmentation probabilities and the average $z$ values are presented in Tables \ref{tbcfp} and \ref{tbc*fp}. Comparing the results in Tables \ref{tbcfp} and \ref{tbc*fp} with those in Tables I, II, III, and IV of ref.\cite{Zheng:2019gnb}, we can see that the fragmentation probability of $g \to B_c(B_c^*)$ is smaller than that of $\bar{b} \to B_c(B_c^*)$ but greater than that of $c \to B_c(B_c^*)$. The fragmentation probabilities for $g \to B_c$ and $g \to B_c^*$ are sensitive to the factorization scale. As the factorization scale increases, the fragmentation probabilities increase significantly. However, the average $z$ values are not sensitive to the factorization scale, they increase slowly as the factorization scale increases. We can also find that the average $z$ value of $D_{g\to B_c}(z,\mu_{F})$ is larger than that of $D_{g\to B_c^*}(z,\mu_{F})$.

\begin{table}[htb]
\centering
\begin{tabular}{c c c c c}
\hline\hline
$\mu_F$(GeV) &  ${\rm P}$($B_c$,ours)~&  ${\rm P}$($B_c$,ref.\cite{Cheung:1993pk})~&  ${\rm P}$($B_c^*$,ours)~ &  ${\rm P}$($B_c^*$,ref.\cite{Cheung:1993pk})~ \\
\hline
15 &  $2.3\times 10^{-5}$ & $1.1\times 10^{-6}$  & $5.6\times 10^{-5}$  &   $1.6\times 10^{-6}$ \\
100 & $4.4\times 10^{-5}$ & $1.9\times 10^{-5}$  & $8.5\times 10^{-5}$  &   $2.8\times 10^{-5}$  \\
800 & $6.7\times 10^{-5}$ & $4.3\times 10^{-5}$  & $1.2\times 10^{-4}$  &   $6.1\times 10^{-5}$  \\
\hline\hline
\end{tabular}
\caption{Comparison of the fragmentation probabilities (for $g \to B_c$ and $g \to B_c^*$) calculated in this work and those given in ref.\cite{Cheung:1993pk}. For consistency with ref.\cite{Cheung:1993pk}, the input parameters for our calculation in this table are taken as the same as those in ref.\cite{Cheung:1993pk}, and the integral interval is taken as $z\in[0.01, 1]$.}
\label{tbcfp-comparison}
\end{table}

It is interesting to give a comparison between our exact results (up to order $\alpha_s^3$) and the approximate results obtained in ref.\cite{Cheung:1993pk}. In Table \ref{tbcfp-comparison}, we present the fragmentation probabilities for $g \to B_c$ and $g \to B_c^*$.  For consistency, the input parameters for our calculation in this table are taken as the same as those in ref.\cite{Cheung:1993pk}, and the integral interval is taken as $z\in[0.01, 1]$. From this table, we can see that our results are significantly different from those in ref.\cite{Cheung:1993pk}, and the relative difference becomes smaller for a larger factorization scale.

For future phenomenological applications, we use polynomials to fit the obtained fragmentation functions. The fragmentation functions can be written as the following form\footnote{The factorization scale $\mu_F$ appearing in the fragmentation functions arises from the operator renormalization of the fragmentation functions, and the $\mu_F$ dependence of the fragmentation functions can be read directly from Eq.(\ref{FFRen}). Therefore, we can divide the fragmentation functions into two parts as shown in Eq.(\ref{eqDzfit}), where the first part depends on $\mu_F$ and the second part is independent of $\mu_F$.}
\begin{eqnarray}
&&D^{\rm NLO}_{g\to B_c(B_c^*)}(z,\mu_{F})\nonumber \\
&&=\frac{\alpha_s}{2\pi}{\rm ln}\frac{\mu_{F}^2}{4M^2}\int_z^1 \Bigg[\sum_{Q=\bar{b},c}P_{Qg}(y) D_{Q\to B_c(B_c^*)}^{\rm LO}(z/y)\Bigg]\frac{dy}{y}+\frac{\alpha_s^3 \vert R_S(0) \vert^2}{M^3} f(z).
\label{eqDzfit}
\end{eqnarray}
For $g \to B_c$, we have
\begin{eqnarray}
f(z)=&&-61.309\,z^8+269.569\,z^7-491.879 \,z^6 + 484.568\, z^5 -280.805 \, z^4\nonumber \\
&& +97.865 \, z^3-20.199\, z^2 +1.888 \, z +0.300.
\label{eqfzBc}
\end{eqnarray}
For $g \to B_c^*$, we have
\begin{eqnarray}
f(z)=&&-300.413\,z^8+1282.128\,z^7-2253.159 \,z^6 + 2105.108\, z^5 -1123.528 \, z^4\nonumber \\
&& +336.215 \, z^3 -45.194\, z^2 -3.034 \, z +1.870 .
\label{eqfzBc*}
\end{eqnarray}

The fragmentation functions for $g \to B_c$ and $g \to B_c^*$ obtained in the present paper and the NLO fragmentation functions for $\bar{b}\to B_c(B_c^*)$ and $c\to B_c(B_c^*)$ obtained in the previous paper \cite{Zheng:2019gnb} provide a full set of fragmentation functions for the $B_c(B_c^*)$ production up to order $\alpha_s^3$. These fragmentation functions can be used as the boundary conditions of the DGLAP evolution equations. For completeness, in addition to the fragmentation functions for $g \to B_c$ and $g \to B_c^*$, we also present the fitting functions to the fragmentation functions for $\bar{b}\to B_c(B_c^*)$ and $c\to B_c(B_c^*)$ in Appendix \ref{Ap.fitfunc}.

\section{Summary}
\label{sum}

In this paper, we have calculated the fragmentation functions for $g\to B_c$ and $g\to B_c^*$, which start at order $\alpha_s^3$. The results obtained in this paper are complementary to the previous calculation on the fragmentation functions for $\bar{b}\to B_c(B_c^*)$ and $c\to B_c(B_c^*)$ up to order $\alpha_s^3$.

There are UV divergences in the phase-space integrals. Dimensional regularization is employed to regularize these UV divergences. However, it is difficult to compute these integrals directly in $d$-dimensional space-time. We adopt the subtraction method to compute these UV divergent integrals. Then the divergent and finite terms are obtained precisely. The UV divergent terms are removed through the renormalization of the operator definition of the fragmentation functions under the $\overline{\rm MS}$ scheme.

The fragmentation functions $D_{g\to B_c}(z,\mu_{F})$ and $D_{g\to B_c^*}(z,\mu_{F})$ under the $\overline{\rm MS}$ factorization scheme are presented as figures and fitting functions. The results show that the fragmentation functions for $g\to B_c$ and $g\to B_c^*$ are sensitive to the factorization scale. As the factorization scale increases, the fragmentation functions increase for $z \in (0,1)$. Thus the fragmentation functions for $g\to B_c$ and $g\to B_c^*$ are important for the $B_c$ production in a very high-energy process. The fragmentation probabilities and the average $z$ values at three typical factorization scales are also presented. The results show that the fragmentation probabilities are sensitive to the factorization scale, while the average $z$ values are insensitive to the factorization scale. Moreover, we found that the fragmentation probabilities for $g\to B_c$ and $g\to B_c^*$ are also positive when $\mu_F=m_b+m_c$ which is below the threshold energy for producing a $B_c$ meson from a gluon. Our results for the fragmentation functions $D_{g\to B_c}(z,\mu_{F})$ and $D_{g\to B_c^*}(z,\mu_{F})$ can be applied to the precision studies of the $B_c$ and $B_c^*$ production at high-energy colliders.

\acknowledgments
This work was supported in part by the Natural Science Foundation of China under Grants No. 11625520, No. 12005028, No. 12175025, No. 12075301, No. 11821505, No. 12047503 and No. 12147102, by the China Postdoctoral Science Foundation under Grant No.2021M693743, by the Fundamental Research Funds for the Central Universities under Grant No.2020CQJQY-Z003, and by the Chongqing Graduate Research and Innovation Foundation under Grant No.ydstd1912. \\


\appendix

\section{Phase-space parametrizations for the subtraction terms}
\label{Ap.para}

To extract the poles in the integrals of the subtraction terms, proper parametrizations to the phase space are required. According to ref.\cite{Zheng:2019gnb}, the differential phase space for a massive particle can be expressed as
\begin{eqnarray}
\frac{d^{d-1}\textbf{p}_i}{(2\pi)^{d-1}2p_i^0}=\frac{(\lambda_i p_i\cdot n-m_i^2)^{-\epsilon}}{4(2\pi)^{3-2\epsilon}}  d\lambda_i d (p_i\cdot n) \,d\Omega_{i\perp},\label{eqa1}
\end{eqnarray}
where
\begin{eqnarray}
\lambda_i=2 k_i\cdot p_i/k_i\cdot n,\label{eqa2}
\end{eqnarray}
$k_i$ is an arbitrary lightlike momentum that is not parallel to $n$, $d\Omega_{i\perp}$ denotes the differential transverse solid angle, and $\Omega_{i\perp}=2\pi^{1-\epsilon}/\Gamma(1-\epsilon)$. Then the differential phase space for the fragmentation function of $g \to (c\bar{b})[n]$ can be expressed as
\begin{eqnarray}
 d\phi_{3}(p_1,p_2,p_3) =&&\frac{(K\cdot n)^{1-2\epsilon}\mu^{4\epsilon}}{16(2\pi)^{5-4\epsilon}}[(1-y)(y-z)]^{-\epsilon}\, \lambda_2^{-\epsilon}\, \lambda_3^{-\epsilon} \left[1-\frac{m_b^2}{\lambda_2(y-z)K\cdot n}\right]^{-\epsilon}  \nonumber \\
&&\times \left[1-\frac{m_c^2}{\lambda_3(1-y)K\cdot n}\right]^{-\epsilon}  dy\, d\lambda_2 \,d\Omega_{2\perp}\,d\lambda_3 \,  d\Omega_{3\perp}, \label{eqa3}
\end{eqnarray}
where the integration over $p_2 \cdot n$ have been performed using the $\delta$ function.

For the $b_1$ and $b_2$ terms in Eq.(\ref{subterm}), we choose the reference lightlike momenta as
\begin{eqnarray}
k_2^{\mu}&=&p_1^{\mu}-\frac{M^2}{2p_1\cdot n}n^{\mu},  \nonumber \\
k_3^{\mu}&=&(p_1+p_2)^{\mu}-\frac{s_1}{2(p_1+p_2)\cdot n}n^{\mu},\label{eqa4}
\end{eqnarray}
then
\begin{eqnarray}
\lambda_2 &=& \frac{1}{zK\cdot n}\left(s_1-\frac{y}{z}M^2-m_b^2\right),\nonumber \\
\lambda_3 &=& \frac{1}{y K\cdot n}\left(s-\frac{s_1}{y}-m_c^2\right).\label{eqa5}
\end{eqnarray}
Changing the variables in Eq.(\ref{eqa3}) from $\lambda_2$ and $\lambda_3$ to $s_1$ and $s$, we obtain
\begin{eqnarray}
&&d\phi_{3}(p_1,p_2,p_3)\nonumber \\
&&=\frac{2^{-2\epsilon}z^{-1+\epsilon}\mu^{4\epsilon}}{(4\pi)^{4-3\epsilon}\Gamma(1-\epsilon)K\cdot n}y^{-1+\epsilon}(1-y)^{-\epsilon}(y-z)^{-\epsilon} \left(s_1-\frac{y M^2}{z}-\frac{y m_b^2}{y-z}\right)^{-\epsilon}  \nonumber \\
&&~~\times\left(s-\frac{s_1}{y}-\frac{m_c^2}{1-y}\right)^{-\epsilon} dy\,ds_1 \, ds \, d\Omega_{3\perp}.  \label{eqa6}
\end{eqnarray}
where the integral over $\Omega_{2\perp}$ is trivial and has been carried out.

For the $b_3$ and $b_4$ terms in Eq.(\ref{subterm}), we choose lightlike momenta as
\begin{eqnarray}
k_2^{\mu}&=&p_1^{\mu}-\frac{M^2}{2p_1\cdot n}n^{\mu},  \nonumber \\
k_3^{\mu}&=&(p_{12}+p_2)^{\mu}-\frac{(p_{12}+p_2)^2}{2(p_{12}+p_2)\cdot n}n^{\mu},\label{eqa7}
\end{eqnarray}
then
\begin{eqnarray}
\lambda_2 &=& \frac{1}{zK\cdot n}\left(s_1-\frac{yM^2}{z}-m_b^2\right),\nonumber \\
\lambda_3 &=& \frac{1}{(y-r_c z) K\cdot n}\left[s_4-\frac{(1-r_c z)}{y-r_c z}(p_{12}+p_2)^2-m_c^2\right].\label{eqa8}
\end{eqnarray}
Changing the variables in Eq.(\ref{eqa3}) from $\lambda_2$ and $\lambda_3$ to $s_4$ and $s_1$, we obtain
\begin{eqnarray}
&&d\phi_{3}(p_1,p_2,p_3)\nonumber \\
&&=\frac{2^{-2\epsilon}[z(y-r_c z)]^{-1+\epsilon}\mu^{4\epsilon}}{(4\pi)^{4-3\epsilon}\Gamma(1-\epsilon)K\cdot n}(1-y)^{-\epsilon}(y-z)^{-\epsilon} \left(s_1-\frac{y\, M^2}{z}-\frac{y\, m_b^2}{y-z}\right)^{-\epsilon}\nonumber \\
&&~~\times  \bigg[s_4-\frac{1-r_c z}{y-r_c z}(p_{12}+p_2)^2-\frac{1-r_c z}{1-y}m_c^2\bigg]^{-\epsilon}dy\, ds_4 \,ds_1 \, d\Omega_{3\perp}.  \label{eqa9}
\end{eqnarray}
where the integration over $\Omega_{2\perp}$ has been performed.

For the $b_5$, $c_5'$ and $d$ terms in Eq.(\ref{subterm}), we choose lightlike momenta as
\begin{eqnarray}
k_2^{\mu}=k_3^{\mu}=p_1^{\mu}-\frac{M^2}{2p_1\cdot n}n^{\mu},\label{eqa10}
\end{eqnarray}
then
\begin{eqnarray}
\lambda_2 &=& \frac{1}{zK\cdot n}\left(s_1-\frac{yM^2}{z}-m_b^2\right),\nonumber \\
\lambda_3 &=& \frac{1}{z K\cdot n}\left[s_2-\frac{(1-y-z)}{z}M^2-m_c^2\right], \label{eqa11}
\end{eqnarray}
Changing the variables in Eq.(\ref{eqa3}) from $\lambda_2$ and $\lambda_3$ to $s_1$ and $s_2$, we obtain
\begin{eqnarray}
&&d\phi_{3}(p_1,p_2,p_3)\nonumber \\
&&=\frac{z^{-2+2\epsilon}\mu^{4\epsilon}}{16(2\pi)^{5-4\epsilon}K\cdot n}(1-y)^{-\epsilon}(y-z)^{-\epsilon}\bigg(s_1-\frac{y M^2}{z}-\frac{y m_b^2}{y-z}\bigg)^{-\epsilon}\nonumber \\
&&~~ \times \left(s_2-\frac{1-y+z}{z}M^2-\frac{1-y+z}{1-y}m_c^2\right)^{-\epsilon} dy\,ds_1 \, ds_2 \, d\Omega_{2\perp}\, d\Omega_{3\perp}.  \label{eqa12}
\end{eqnarray}

For the $c_1$ and $c_2$ terms in Eq.(\ref{subterm}), we choose lightlike momenta as
\begin{eqnarray}
k_2^{\mu}&=&(p_1+p_3)^{\mu}-\frac{(p_1+p_3)^2}{2(p_1+p_3)\cdot n}n^{\mu}, \nonumber \\
k_3^{\mu}&=&p_1^{\mu}-\frac{M^2}{2p_1\cdot n}n^{\mu}, \label{eqa13}
\end{eqnarray}
then
\begin{eqnarray}
\lambda_2 &=& \frac{1}{(1-y+z)K\cdot n}\left(s-\frac{s_2}{1-y+z}-m_b^2\right),\nonumber \\
\lambda_3 &=& \frac{1}{z K\cdot n}\left(s_2-\frac{1-y+z}{z}M^2-m_c^2\right),\label{eqa14}
\end{eqnarray}
Changing the variables in Eq.(\ref{eqa3}) from $\lambda_2$ and $\lambda_3$ to $s_2$ and $s$, we obtain
\begin{eqnarray}
&&d\phi_{3}(p_1,p_2,p_3)\nonumber \\
&&=\frac{2^{-2\epsilon}z^{-1+\epsilon}(1-y+z)^{-1+\epsilon}\mu^{4\epsilon}}{(4\pi)^{4-3\epsilon}\Gamma(1-\epsilon)K\cdot n}(1-y)^{-\epsilon}(y-z)^{-\epsilon} \nonumber \\
&&~~\times \left(s_2-\frac{1-y+z}{z}M^2-\frac{1-y+z }{1-y}m_c^2\right)^{-\epsilon} \bigg(s-\frac{s_2}{1-y+z} \nonumber \\
&&~~-\frac{m_b^2}{y-z}\bigg)^{-\epsilon} dy\,ds_2 \, ds \, d\Omega_{2\perp},  \label{eqa15}
\end{eqnarray}
where the integration over $\Omega_{3\perp}$ has been performed.

For the $c_3$ and $c_4$ terms in Eq.(\ref{subterm}), we choose lightlike momenta as
\begin{eqnarray}
k_2^{\mu}&=&(p_{11}+p_3)^{\mu}-\frac{(p_{11}+p_3)^2}{2(p_{11}+p_3)\cdot n}n^{\mu}, \nonumber \\
k_3^{\mu}&=&p_1^{\mu}-\frac{M^2}{2p_1\cdot n}n^{\mu}, \label{eqa16}
\end{eqnarray}
then
\begin{eqnarray}
\lambda_2 =&& \frac{1}{(1-y+r_c z)K\cdot n}\bigg[s_3-\frac{1-r_b z}{1-y+r_c z} (p_{11}+p_3)^2-m_b^2\bigg],\nonumber \\
\lambda_3 =&& \frac{1}{z K\cdot n}\left(s_2-\frac{1-y+z}{z}M^2-m_c^2\right).\label{eqa17}
\end{eqnarray}
Changing the variables in Eq.(\ref{eqa3}) from $\lambda_2$ and $\lambda_3$ to $s_2$ and $s_3$, we obtain
\begin{eqnarray}
&&d\phi_{3}(p_1,p_2,p_3)\nonumber \\
&&=\frac{2^{-2\epsilon}z^{-1+\epsilon}(1-y+r_c z)^{-1+\epsilon}\mu^{4\epsilon}}{(4\pi)^{4-3\epsilon}\Gamma(1-\epsilon)K\cdot n}(1-y)^{-\epsilon}(y-z)^{-\epsilon} \nonumber \\
&&~~\times \left(s_2-\frac{1-y+z}{z}M^2-\frac{1-y+z }{1-y}m_c^2\right)^{-\epsilon}\nonumber \\
&&~~\times  \bigg[s_3-\frac{1-r_b z}{1-y+r_c z}(p_{11}+p_3)^2-\frac{1-r_b z }{y-z}m_b^2\bigg]^{-\epsilon} \nonumber \\
&&~~ dy\,ds_2 \, ds_3 \, d\Omega_{2\perp}.  \label{eqa18}
\end{eqnarray}
where the integration over $\Omega_{3\perp}$ has been performed.

\section{Integrals of the subtraction terms}
\label{Ap.int}

Having the parametrizations derived in Appendix \ref{Ap.para} for the phase space, the integrals of the subtraction terms can be carried out in $d$-dimensional space time.

To integrate the $b_1$ and $b_2$ terms in Eq.(\ref{subterm}) over the phase space, we use the parametrization in Eq.(\ref{eqa6}). The expression of Eq.(\ref{eqa6}) can be further decomposed as
\begin{eqnarray}
&& N_{\rm CS}\,d\phi_3(p_1,p_2,p_3) =N_{q}(p_1,p_2)d\phi_2(p_1,p_2)d\phi^{(3)}(p_1,p_2,p_3),\label{eqb1}
\end{eqnarray}
where $N_{q}(p_1,p_2)$ is defined as
\begin{eqnarray}
N_{q}(p_1,p_2)=\frac{(z/y)^{1-2\epsilon}}{8\pi N_c},\label{eqb2}
\end{eqnarray}
and $d\phi_{2}(p_1,p_2)$ is defined as
\begin{eqnarray}
d\phi_{2}(p_1,p_2)=&&\frac{z^{-1+\epsilon}(y-z)^{-\epsilon}\mu^{2\epsilon}}{2(4\pi)^{1-\epsilon}\Gamma(1-\epsilon)K\cdot n} \left(s_1-\frac{y}{z}M^2-\frac{y}{y-z}m_b^2\right)^{-\epsilon}ds_1.\label{eqb3}
\end{eqnarray}
Then the expression of $d\phi^{(3)}(p_1,p_2,p_3)$ can be written down, i.e.,
\begin{eqnarray}
d\phi^{(3)}(p_1,p_2,p_3)=&&\frac{N_c \,[y(1-y)]^{-\epsilon}\mu^{2\epsilon}}{(N_c^2-1)(2-2\epsilon)(2\pi)^{3-2\epsilon}K\cdot n} \left(s-\frac{s_1}{y}-\frac{m_c^2}{1-y}\right)^{-\epsilon}ds\, dy\, d\Omega_{3\perp}.\label{eqb4}
\end{eqnarray}
The range of $y$ is from $z$ to 1, the range of $s_1$ is from $[y\,M^2/z+y\, m_b^2/(y-z)]$ to $\infty$, and the range of $s$ is from $[s_1/y+m_c^2/(1-y)]$ to $\infty$.

With Eqs.(\ref{eqb1})-(\ref{eqb4}), we obtain
\begin{eqnarray}
&&N_{\rm CS}\int d\phi_{3}(p_1,p_2,p_3)\frac{b_1(s_1,z,y)}{s}\nonumber \\
&&=\frac{4\,N_c \,\Gamma(\epsilon)\,\mu^{2\epsilon}}{(N_c^2-1)(2-2\epsilon)(4\pi)^{2-\epsilon}K\cdot n}\int_z^1 dy [y(1-y)]^{-\epsilon} \nonumber \\
&& ~~~ \times \int N_q d\phi_{2}(p_1,p_2)b_1(s_1,z,y) \left(\frac{s_1}{y}+\frac{m_c^2}{1-y}\right)^{-\epsilon},\label{eqb5}
\end{eqnarray}
where $N_qd\phi_{2}(p_1,p_2)$ is the abbreviation of $N_{q}(p_1,p_2)d\phi_{2}(p_1,p_2)$.

For the $b_1$ term, the integration over $\Omega_{3\perp}$ is trivial, while for the the $b_2$ term, the integration  over $\Omega_{3\perp}$ is nontrivial. To integrate the $b_2$ term over the phase space, we first integrate $p_3^{\mu}$ over $\Omega_{3\perp}$. According to Lorentz invariance, the integral over $\Omega_{3\perp}$ has the following form:
\begin{eqnarray}
\int p_3^{\mu} d\Omega_{3\perp}=A n^{\mu}+B(p_1+p_2)^{\mu}.\label{eqb6}
\end{eqnarray}
To determine the coefficients $A$ and $B$, we contract both sides of Eq.(\ref{eqb6}) with $n^{\mu}$ [and contract with $(p_1+p_2)^{\mu}$]. We obtain
\begin{eqnarray}
A&=&\frac{\Omega_{\perp}}{2yK\cdot n}\left(s-\frac{2-y}{y}s_1-m_c^2\right),\label{eqb7}\\
B&=&\frac{1-y}{y}\Omega_{\perp},\label{eqb8}
\end{eqnarray}
where $\Omega_{\perp}=2\pi^{1-\epsilon}/\Gamma(1-\epsilon)$. Contracting both sides of Eq.(\ref{eqb6}) with $p_1^{\mu}$, we obtain
\begin{eqnarray}
\int p_1\cdot p_3\, d\Omega_{3\perp}=\frac{z\,s\,\Omega_{\perp}}{2\,y}+f_1\,\Omega_{\perp},\label{eqb9}
\end{eqnarray}
where $f_1$ has been given in Eq.(\ref{subterm-f1}). This is equivalent to
\begin{eqnarray}
\int (p_1\cdot p_3-f_1)\, d\Omega_{3\perp}=\frac{z\,s\,\Omega_{\perp}}{2\,y}.\label{eqb10}
\end{eqnarray}
Then the integration over phase space of $b_2$ term can be carried out. We have
\begin{eqnarray}
&&N_{\rm CS}\int d\phi_{3}(p_1,p_2,p_3)\frac{b_2(s_1,z,y)(p_1\cdot p_3-f_1)}{s^2} \nonumber \\
&&=\frac{2\,N_c\,z \,\Gamma(\epsilon)\,\mu^{2\epsilon}}{(N_c^2-1)(2-2\epsilon)(4\pi)^{2-\epsilon}K\cdot n}\int_z^1 dy y^{-1-\epsilon}(1-y)^{-\epsilon} \nonumber \\
&& ~~~ \times \int N_q d\phi_{2}(p_1,p_2)b_2(s_1,z,y) \left(\frac{s_1}{y}+\frac{m_c^2}{1-y}\right)^{-\epsilon}.\label{eqb11}
\end{eqnarray}

To integrate the $b_3$ and $b_4$ terms in Eq.(\ref{subterm}) over the phase space, we use the parametrization of Eq.(\ref{eqa9}). The parametrization of Eq.(\ref{eqa9}) can also be decomposed as the form of Eq.(\ref{eqb1}), where
\begin{eqnarray}
d\phi^{(3)}(p_1,p_2,p_3)=&&\frac{N_c \,y^{1-2\epsilon}(1-y)^{-\epsilon}(y-r_c z)^{-1+\epsilon}\mu^{2\epsilon}}{(N_c^2-1)(2-2\epsilon)(2\pi)^{3-2\epsilon}K\cdot n}\nonumber \\
&&\times \bigg[s_4-\frac{1-r_c z}{y-r_c z}(p_{12}+p_2)^2 -\frac{(1-r_c z)m_c^2}{1-y}\bigg]^{-\epsilon}ds_4\, dy\, d\Omega_{3\perp},\label{eqb12}
\end{eqnarray}
where $(p_{12}+p_2)^2=r_b(s_1-m_c^2)$. The ranges of $y$ and $s_1$ are the same as those in Eqs.(\ref{eqb3})-(\ref{eqb4}), while the range of $s_4$ is from $(1-r_c z)[r_b(s_1-m_c^2)/(y-r_c z)+m_c^2/(1-y)]$ to $\infty$.

The integration of the $b_3$ term over $\Omega_{3\perp}$ and $s_4$ can be carried out, and we obtain
\begin{eqnarray}
&&N_{\rm CS}\int d\phi_{3}(p_1,p_2,p_3)\frac{b_3(s_1,z,y)}{s_4-m_c^2}\nonumber \\
&&=\frac{4\,N_c \,\Gamma(\epsilon)\,\mu^{2\epsilon}}{(N_c^2-1)(2-2\epsilon)(4\pi)^{2-\epsilon}K\cdot n}\int_z^1 dy y^{1-2\epsilon}(1-y)^{-\epsilon}  (y-r_c z)^{-1+\epsilon}\nonumber \\
&& ~~~ \times\int N_q d\phi_{2}(p_1,p_2)b_3(s_1,z,y) \bigg[\frac{(1-r_c z)}{y-r_c z}(p_{12}+p_2)^2+\frac{(y-r_c z)m_c^2}{1-y}\bigg]^{-\epsilon}.\label{eqb13}
\end{eqnarray}
Applying the method presented in Eqs.(\ref{eqb6})-(\ref{eqb11}) to the integration of the $b_4$ term, we obtain
\begin{eqnarray}
&&N_{\rm CS}\int d\phi_{3}(p_1,p_2,p_3)\frac{b_4(s_1,z,y)(p_1\cdot p_3-f_2)}{(s_4-m_c^2)^2} \nonumber \\
&&=\frac{2\,N_c \,z\, \Gamma(\epsilon)\,\mu^{2\epsilon}}{(N_c^2-1)(2-2\epsilon)(4\pi)^{2-\epsilon}K\cdot n}\int_z^1 dy y^{1-2\epsilon}(1-y)^{-\epsilon}(y-r_c z)^{-2+\epsilon} \nonumber \\
&& ~~~ \times \int N_q d\phi_{2}(p_1,p_2)b_4(s_1,z,y) \bigg[\frac{(1-r_c z)}{y-r_c z}(p_{12}+p_2)^2+\frac{(y-r_c z)m_c^2}{1-y}\bigg]^{-\epsilon}.\label{eqb14}
\end{eqnarray}

To integrate the $b_5$ and $d$ terms in Eq.(\ref{subterm}) over the phase space, we adopt the parametrization in Eq.(\ref{eqa12}). The differential phase space given in Eq.(\ref{eqa12}) can also be expressed as the form of Eq.(\ref{eqb1}), and
\begin{eqnarray}
d\phi^{(3)}(p_1,p_2,p_3)=&&\frac{N_c \,z^{-1+\epsilon}y^{1-2\epsilon}(1-y)^{-\epsilon}\mu^{2\epsilon}}{(N_c^2-1)(2-2\epsilon)(2\pi)^{3-2\epsilon}K\cdot n} \bigg[s_2-\frac{(1-y+z)M^2}{z} \nonumber \\
&& -\frac{(1-y+z)m_c^2}{1-y} \bigg]^{-\epsilon} ds_2\, dy\, d\Omega_{3\perp}.\label{eqb15}
\end{eqnarray}
The range of $s_2$ is from $(1-y+ z)[M^2/z+m_c^2/(1-y)]$ to $\infty$. Then we obtain
\begin{eqnarray}
&&N_{\rm CS}\int d\phi_{3}(p_1,p_2,p_3)\frac{b_5(s_1,z,y)}{s_2-m_b^2}\nonumber \\
&&=\frac{4\,N_c \,\Gamma(\epsilon)\,z^{-1+\epsilon}\,\mu^{2\epsilon}}{(N_c^2-1)(2-2\epsilon)(4\pi)^{2-\epsilon}K\cdot n}\int_z^1 dy y^{1-2\epsilon}(1-y)^{-\epsilon} (1-y+ z)^{-\epsilon} \nonumber \\
&& ~~~ \times \int N_q d\phi_{2}(p_1,p_2)b_5(s_1,z,y) \bigg[\frac{M^2}{z} +\frac{m_c^2}{1-y}-\frac{m_b^2}{1-y+z}\bigg]^{-\epsilon},\label{eqb16}
\end{eqnarray}
and
\begin{eqnarray}
&&N_{\rm CS}\int d\phi_{3}(p_1,p_2,p_3)\frac{d(z,y)(p_2\cdot p_3-f_3)}{(s_1-m_c)^2(s_2-m_b^2)^2}\nonumber \\
&&=\frac{2\,N_c \,\Gamma(\epsilon)\,z^{-2+\epsilon}\,\mu^{2\epsilon}}{(N_c^2-1)(2-2\epsilon)(4\pi)^{2-\epsilon}K\cdot n}\int_z^1 dy y^{1-2\epsilon}(1-y)^{-\epsilon} (1-y+ z)^{-\epsilon}(y-z)\nonumber \\
&& ~~~ \times \int N_q d\phi_{2}(p_1,p_2)\frac{d(z,y)}{(s_1-m_c)^2} \bigg[\frac{M^2}{z}+\frac{m_c^2}{1-y}-\frac{m_b^2}{1-y+z}\bigg]^{-\epsilon}.\label{eqb17}
\end{eqnarray}

To integrate the $c_1$ and $c_2$ terms in Eq.(\ref{subterm}) over the phase space, we adopt the parametrization in Eq.(\ref{eqa15}). The differential phase space given in Eq.(\ref{eqa15}) can be expressed as
\begin{eqnarray}
&& N_{\rm CS}\,d\phi_3(p_1,p_2,p_3) =N_{q}(p_1,p_3)d\phi_2(p_1,p_3)d\phi^{(2)}(p_1,p_2,p_3),\label{eqb18}
\end{eqnarray}
where
\begin{eqnarray}
N_{q}(p_1,p_3)=\frac{[z/(1-y+z)]^{1-2\epsilon}}{8\pi N_c},\label{eqb19}
\end{eqnarray}
and
\begin{eqnarray}
d\phi_{2}(p_1,p_3)=&&\frac{z^{-1+\epsilon}(1-y)^{-\epsilon}\mu^{2\epsilon}}{2(4\pi)^{1-\epsilon}\Gamma(1-\epsilon)K\cdot n}\bigg(s_2 -\frac{1-y+z}{z}M^2-\frac{1-y+z}{1-y}m_c^2\bigg)^{-\epsilon}ds_2.\nonumber \\ \label{eqb20}
\end{eqnarray}
Then we obtain
\begin{eqnarray}
d\phi^{(2)}(p_1,p_2,p_3)=&&\frac{N_c \,[(1-y+z)(y-z)]^{-\epsilon}\mu^{2\epsilon}}{(N_c^2-1)(2-2\epsilon)(2\pi)^{3-2\epsilon}K\cdot n}  \left[s-\frac{s_2}{1-y+z}-\frac{m_b^2}{y-z} \right]^{-\epsilon}\nonumber \\
&& \times ds\, dy\, d\Omega_{2\perp}.\label{eqb21}
\end{eqnarray}
The range of $y$ is from $z$ to 1, the range of $s_2$ is from $(1-y+ z)[M^2/z+m_c^2/(1-y)]$ to $\infty$ and the range of $s$ is from $[s_2/(1-y+z)+m_b^2/(y-z)]$ to $\infty$.

The integration of the $c_1$ and $c_2$ terms over $\Omega_{2\perp}$ and $s$ can be carried out easily, and we obtain
\begin{eqnarray}
&&N_{\rm CS}\int d\phi_{3}(p_1,p_2,p_3)\frac{c_1(s_2,z,y)}{s}\nonumber \\
&&=\frac{4\,N_c \,\Gamma(\epsilon)\,\mu^{2\epsilon}}{(N_c^2-1)(2-2\epsilon)(4\pi)^{2-\epsilon}K\cdot n}\int_z^1 dy (y-z)^{-\epsilon}(1-y+ z)^{-\epsilon} \nonumber \\
&& ~~~ \times \int N_q d\phi_{2}(p_1,p_3)c_1(s_2,z,y) \bigg(\frac{s_2}{1-y+z} +\frac{m_b^2}{y-z}\bigg)^{-\epsilon},\label{eqb22}
\end{eqnarray}
and
\begin{eqnarray}
&&N_{\rm CS}\int d\phi_{3}(p_1,p_2,p_3)\frac{c_2(s_2,z,y)(p_1\cdot p_2-g_1)}{s^2}\nonumber \\
&&=\frac{2\,N_c \,z\,\Gamma(\epsilon)\,\mu^{2\epsilon}}{(N_c^2-1)(2-2\epsilon)(4\pi)^{2-\epsilon}K\cdot n}\int_z^1 dy (y-z)^{-\epsilon}(1-y+ z)^{-1-\epsilon} \nonumber \\
&& ~~~ \times \int N_q d\phi_{2}(p_1,p_3)c_2(s_2,z,y)  \left(\frac{s_2}{1-y+z} +\frac{m_b^2}{y-z}\right)^{-\epsilon},\label{eqb23}
\end{eqnarray}
where $N_q d\phi_{2}(p_1,p_3) \equiv N_q(p_1,p_3) d\phi_{2}(p_1,p_3)$.

For the $c_3$ and $c_4$ terms in Eq.(\ref{subterm}), we use the parametrization in Eq.(\ref{eqa18}). The differential phase space given in Eq.(\ref{eqa18}) can also be expressed as the form of Eq.(\ref{eqb18}), and
\begin{eqnarray}
d\phi^{(2)}(p_1,p_2,p_3)=&&\frac{N_c \,(y-z)^{-\epsilon}(1-y+z)^{1-2\epsilon}\mu^{2\epsilon}}{(N_c^2-1)(2-2\epsilon)(2\pi)^{3-2\epsilon}K\cdot n}(1-y+r_c z)^{-1+\epsilon} \nonumber \\
&&\times  \bigg[s_3-\frac{(1-r_b z)r_c}{1-y+r_c z} (s_2-m_b^2)-\frac{1-r_b z}{y-z} m_b^2\bigg]^{-\epsilon}ds_3\, dy\, d\Omega_{2\perp}.\label{eqb24}
\end{eqnarray}
The ranges of $y$ and $s_2$ are the same as those of Eq.(\ref{eqb20}), and the range of $s_3$ is from $(1-r_b z)[r_c(s_2-m_b^2)/(1-y+r_c z)+m_b^2/(y-z)]$ to $\infty$.

Performing the integration of the $c_3$ and $c_4$ terms over $\Omega_{2\perp}$ and $s_3$, we obtain
\begin{eqnarray}
&&N_{\rm CS}\int d\phi_{3}(p_1,p_2,p_3)\frac{c_3(s_2,z,y)}{s_3-m_b^2}\nonumber \\
&&=\frac{4\,N_c \,\Gamma(\epsilon)\,\mu^{2\epsilon}}{(N_c^2-1)(2-2\epsilon)(4\pi)^{2-\epsilon}K\cdot n}\int_z^1 dy (y-z)^{-\epsilon} (1-y+ z)^{1-2\epsilon}(1-y+r_c z)^{-1+\epsilon}\nonumber \\
&& ~~~ \times \int N_q d\phi_{2}(p_1,p_3)c_3(s_2,z,y)\bigg[\frac{(1-r_b z)r_c}{1-y+r_c z}(s_2-m_b^2)  +\frac{1-y+r_c z}{y-z} m_b^2\bigg]^{-\epsilon},\label{eqb25}
\end{eqnarray}
and
\begin{eqnarray}
&&N_{\rm CS}\int d\phi_{3}(p_1,p_2,p_3)\frac{c_4(s_2,z,y)(p_1\cdot p_2-g_2)}{(s_3-m_b^2)^2}\nonumber \\
&&=\frac{2\,N_c \,z\,\Gamma(\epsilon)\,\mu^{2\epsilon}}{(N_c^2-1)(2-2\epsilon)(4\pi)^{2-\epsilon}K\cdot n}\int_z^1 dy (y-z)^{-\epsilon}(1-y+ z)^{1-2\epsilon}(1-y+r_c z)^{-2+\epsilon} \nonumber \\
&& ~~~ \times \int N_q d\phi_{2}(p_1,p_3) c_4(s_2,z,y) \bigg[\frac{(1-r_b z)r_c}{1-y+r_c z}(s_2-m_b^2) +\frac{1-y+r_c z}{y-z} m_b^2\bigg]^{-\epsilon}.\label{eqb26}
\end{eqnarray}

For the $c_5'$ term in Eq.(\ref{subterm}), we use the parametrization in Eq.(\ref{eqa12}). The differential phase space given in Eq.(\ref{eqa12}) can also be expressed as the form of Eq.(\ref{eqb17}), and
\begin{eqnarray}
d\phi^{(2)}(p_1,p_2,p_3)=&&\frac{N_c \,z^{-1+\epsilon}\,(y-z)^{-\epsilon}(1-y+z)^{1-2\epsilon}\mu^{2\epsilon}}{(N_c^2-1)(2-2\epsilon)(2\pi)^{3-2\epsilon}K\cdot n} \nonumber \\
&&\times  \bigg(s_1-\frac{y\, M^2}{z} -\frac{y\,m_b^2}{y-z} \bigg)^{-\epsilon}ds_1\, dy\, d\Omega_{2\perp}.\label{eqb27}
\end{eqnarray}
The ranges of $y$ and $s_2$ are the same as those of Eq.(\ref{eqb20}), and the range of $s_1$ is from $[y\,M^2/z+y\,m_b^2/(y-z)]$ to $\infty$.
Performing the integration of the $c_5'$ term over $\Omega_{2\perp}$ and $s_1$, we obtain
\begin{eqnarray}
&&N_{\rm CS}\int d\phi_{3}(p_1,p_2,p_3)\frac{c_5'(s_2,z,y)}{s_1-m_c^2}\nonumber \\
&&=\frac{4\,N_c \,\Gamma(\epsilon)\,z^{-1+\epsilon}\,\mu^{2\epsilon}}{(N_c^2-1)(2-2\epsilon)(4\pi)^{2-\epsilon}K\cdot n}\int_z^1 dy [y(y-z)]^{-\epsilon} (1-y+ z)^{1-2\epsilon}\nonumber \\
&& ~~~ \times \int N_q d\phi_{2}(p_1,p_3)c_5'(s_2,z,y)  \bigg[\frac{M^2}{z} +\frac{m_b^2}{y-z}- \frac{m_c^2}{y}\bigg]^{-\epsilon}.\label{eqb28}
\end{eqnarray}

The remaining integrals in this appendix does not generate poles in $\epsilon$ any more. Therefore, we can expand $\epsilon$ before performing the remaining integrations.

\section{Fitting functions for the fragmentation functions of $\bar{b} \to B_c(B_c^*)$ and $c \to B_c(B_c^*)$}
\label{Ap.fitfunc}

The NLO fragmentation functions for $\bar{b} \to B_c(B_c^*)$ and $c \to B_c(B_c^*)$ have been obtained in our previous work \cite{Zheng:2019gnb}. In this appendix, we present the fitting functions to those fragmentation functions. The fragmentation function for $\bar{b} \to B_c(B_c^*)$ can be written as
\begin{eqnarray}
D^{\rm NLO}_{\bar{b}\to B_c(B_c^*)}(z,\mu_{F})=&& D^{\rm LO}_{\bar{b}\to B_c(B_c^*)}(z)\left(1+\frac{\alpha_s(\mu_R)}{2\pi}\beta_0 {\rm ln}\frac{\mu_R^2}{4m_c^2}\right)+\frac{\alpha_s(\mu_R)}{2\pi}{\rm ln}\frac{\mu_{F}^2}{(m_b+2m_c)^2}\nonumber \\
&&\times \int_z^1 \frac{dy}{y}\Big[P_{\bar{b}\bar{b}}(y) D_{\bar{b}\to B_c(B_c^*)}^{\rm LO}(z/y)\Big]+\frac{\alpha_s(\mu_R)^3 \vert R_S(0) \vert^2}{M^3} g(z),
\label{eqDzfit2}
\end{eqnarray}
where $\beta_0=11-2n_f/3$, $n_f$ is the number of active quark flavors, and the splitting function
\begin{eqnarray}
P_{\bar{b}\bar{b}}(y)=\frac{4}{3}\left[\frac{1+y^2}{(1-y)_+} +\frac{3}{2}\delta(1-y)\right].
\end{eqnarray}
For $\bar{b} \to B_c$, we have
\begin{eqnarray}
g(z)=&&-1435171.329\,z^{14}+9392402.604\,z^{13}- 27433834.065\,z^{12} +47172402.976 \,z^{11} \nonumber\\
&& - 53031470.170\,z^{10} + 40932167.342 \,z^9 - 22163212.706 \,z^8 + 8450547.448 \,z^7 \nonumber\\
&& - 2243748.824 \,z^6 +
  403596.737 \,z^5 - 46759.152 \,z^4 + 3182.230 \,z^3 - 102.993 \,z^2 \nonumber\\
&& + 0.413 \,z - 0.502.
\label{eqgzBc}
\end{eqnarray}
For $\bar{b} \to B_c^*$, we have
\begin{eqnarray}
g(z)=&&-4688679.390 \,z^{14} + 30865734.204 \,z^{13} - 90699028.409 \,z^{12} +
 156934867.732 \,z^{11} \nonumber\\
&& - 177595170.016 \,z^{10} + 138055174.638 \,z^9 -
 75343441.896 \,z^8 + 28988875.618 \,z^7\nonumber\\
&&  - 7781479.122 \,z^6 +
 1419561.311 \,z^5 - 167804.824 \,z^4 + 11802.930 \,z^3 - 413.741 \,z^2 \nonumber\\
&& + 1.364 \,z - 0.360.
\label{eqgzBc*}
\end{eqnarray}

The NLO fragmentation function for $c \to B_c(B_c^*)$ can be written as
\begin{eqnarray}
D^{\rm NLO}_{c\to B_c(B_c^*)}(z,\mu_{F})=&& D^{\rm LO}_{c\to B_c(B_c^*)}(z)\left(1+\frac{\alpha_s(\mu_R)}{2\pi}\beta_0 {\rm ln}\frac{\mu_R^2}{4m_b^2}\right)+\frac{\alpha_s(\mu_R)}{2\pi}{\rm ln}\frac{\mu_{F}^2}{(2m_b+m_c)^2}\nonumber \\
&&\times \int_z^1 \frac{dy}{y}\Big[P_{cc}(y) D_{c\to B_c(B_c^*)}^{\rm LO}(z/y)\Big]+\frac{\alpha_s(\mu_R)^3 \vert R_S(0) \vert^2}{M^3} h(z),
\label{eqDzfit3}
\end{eqnarray}
where $P_{cc}(y)=P_{\bar{b}\bar{b}}(y)$. For $c \to B_c$, we have
\begin{eqnarray}
h(z)=&& 1.3747 \,z^8 - 2.7415 \,z^7 - 0.1967 \,z^6 + 5.8231 \,z^5 - 6.7841 \,z^4 + 2.5408 \,z^3 \nonumber \\
&&  - 0.3821 \,z^2+ 0.3669 \,z - 0.0017.
\label{eqhzBc}
\end{eqnarray}
For $c \to B_c^*$, we have
\begin{eqnarray}
h(z)=&& 4.7301 \,z^8 - 19.9862 \,z^7 + 35.4333 \,z^6 - 32.1254 \,z^5 + 14.1958 \,z^4 - 2.2313 \,z^3  \nonumber \\
&&- 0.1102 \,z^2 + 0.0655 \,z + 0.0279.
\label{eqhzBc*}
\end{eqnarray}
More details of the calculation on these fragmentation functions can be found in ref.\cite{Zheng:2019gnb}.

\providecommand{\href}[2]{#2}\begingroup\raggedright

\end{document}